\documentclass[11pt,a4paper]{article} 
\usepackage{style}

\usepackage{amsmath,amssymb,epsfig,bm,amsfonts,yfonts,bbm,mathrsfs,hyperref}
\usepackage{dsfont,mathtools,graphicx,empheq,color,tikz,stmaryrd,gensymb,braket}
\usepackage{dynkin-diagrams}

\definecolor{rot}{rgb}{0.7,0,0}
\definecolor{gruen}{rgb}{0,0.7,0}
\definecolor{blau}{rgb}{0,0,0.6}

\title{Exact flow equation for the divergence functional}

\author{Stefan Floerchinger}

\affiliation{Theoretisch-Physikalisches Institut, Max-Wien Platz 1, 07743 Jena, Germany}

\emailAdd{stefan.floerchinger@uni-jena.de}

\abstract{An exact functional renormalization group flow equation is derived for the  divergence functional which is a generalization of the Kullback-Leibler divergence to quantum field theories in the Euclidean domain. It compares distributions with different sources and field expectation values. The renormalization group flow for a regularized version of this functional connects two limits: one where the functional is known in terms of the microscopic action or probability distribution, and the other where all fluctuations are taken into account. In the latter limit one can obtain full correlation functions from functional derivatives of the divergence functional. The flow equation provides a possiblity to determine this functional non-perturbatively.}

\begin{document}

\maketitle

\section{Introduction}

The functional renormalization group is a method widely employed in quantum field theory and statistical physics, see refs.\ \cite{morrisElementsContinuousRenormalization1998, bagnulsExactRenormalizationGroup2001, salmhoferFermionicRenormalizationGroup2001, bergesNonPerturbativeRenormalizationFlow2002,polonyiLecturesFunctionalRenormalization2003,delamotteIntroductionNonperturbativeRenormalization2012,giesIntroductionFunctionalRG2012,pawlowskiAspectsFunctionalRenormalisation2007, reuterFunctionalRenormalizationGroup2007, metznerFunctionalRenormalizationGroup2012, rostenFundamentalsExactRenormalization2012, camachoReviewRecentDevelopments2022,kopietzIntroductionFunctionalRenormalization2010, dupuisNonperturbativeFunctionalRenormalization2021,wipfStatisticalApproachQuantum2021} for reviews. It allows to connect the microscopic description of a quantum field theory in terms of its microscopic action to the macroscopic regime where the theory is described by a generating functional for correlation functions such as the Schwinger functional or its Legendre transform, the quantum effective action. The method is based on exact renormalization group flow equations, the Polchinski equation for a modified version of the Schinger functional \cite{Polchinski:1983gv}, or the Wetterich equation for a modified version of the quantum effective action \cite{Wetterich:1992yh}. For an extension with a scale-dependent Hubbard-Stratonovich transformation see ref.\ \cite{FloerchingerExactFlowEquation2009}. Predecessors of these equations are the equation of Wegner and Houghton \cite{wegnerRenormalizationGroupEquation1973} or the Callan-Symmanzik equation \cite{callanBrokenScaleInvariance1970, symanzikSmallDistanceBehaviour1970, symanzikSmalldistancebehaviourAnalysisWilson1971}. 

In the present article we derive a similar renormalization group flow equation for the divergence functional, which is a functional version of the Kullback-Leibler divergence \cite{kullbackInformationSufficiency1951, kullbackInformationTheoryStatistics1997}. Formally it corresponds to the relative information entropy between probability distributions characterized by two different values of the source or field expectation value, respectively. Interestingly, this functional can also be seen as a generating functional for correlation functions, similar as the quantum effective action \cite{FloerchingerInformationGeometryEuclidean2023}. 

The Kullback-Leibler divergence plays an important role in probability and information theory \cite{kullbackInformationTheoryStatistics1997,coverElementsInformationTheory2006}. Essentially it describes how well one distribution can be differentiated from another. It governs also the probability for distributions of field configuations after asymptotically many drawings through Sanov's theorem \cite{coverElementsInformationTheory2006}. This object also plays an important role in information geometry \cite{amariInformationGeometryIts2016,ayInformationGeometry2017}. Geometric quantities like the Fisher information metric or the two dual Amari-Chentsov connections can be derived from it, see also \cite{FloerchingerInformationGeometryEuclidean2023}.

The present work can be seen as a continuation of previous works aiming to explore the connections between geometry, information theory and the renormalization group \cite{diosiMetricizationThermodynamicstateSpace1984,gaiteFieldTheoryEntropy1996,dolanRenormalisationGroupFlow1998,
brodySymmetryRealspaceRenormalisation1998, dolanRenormalizationGroupFlow1999,preskillQuantumInformationPhysics2000, casiniTheoremEntanglementEntropy2007, apenkoInformationTheoryRenormalization2012, benyInformationgeometricApproachRenormalization2015, benyRenormalizationGroupStatistical2015, fowlerMisanthropicEntropyRenormalization2022,koenigsteinNumericalFluidDynamics2022,erdmengerQuantifyingInformationFlows2022}.
Recently also connections between the renormalization group and optimal transport have been explored \cite{cotlerRenormalizationGroupFlow2023, bermanInverseExactRenormalization2022}.

\section{Modified exponential class of probability densities}

We consider a class of probability densities with respect to a measure $D\phi$ for random variables or fields $\phi_\alpha$,
\begin{equation}
  p_k[\phi, J] = \exp\left( -S[\phi] - \frac{1}{2} R_k^{\alpha\beta} \phi_\alpha \phi_\beta + J^\alpha \phi_\alpha - W_k[J]\right),
  \label{eq:functionalProbDensitySourceCoordinates}
\end{equation}
with the Schwinger functional $W_k[J]$ defined such that the probability density is properly normalized,
\begin{equation}
  e^{W_k[J]} = \int D \phi \exp\left( -S[\phi] - \frac{1}{2} R_k^{\alpha\beta} \phi_\alpha \phi_\beta + J^\alpha \phi_\alpha \right).
  \label{eq:modifiedSchwingerFunctional}
\end{equation}
We use an abstract index notation where $\alpha$ combines continuous indices like spatial position and abstract indices like field components, for axample $\alpha=(x,j)$ and $\phi_\alpha = \phi_j(x)$. Einsteins summation convention implies a sum over discrete indices and an integral over continuous indices. It is usually clear from the context how expressions in this shorthand notation can be made more concrete.

The sources $J^\alpha$ are also fields, while $R_k^{\alpha\beta}$ is considered as a matrix parameterized by a flow parameter $k$. We are interested in deriving differential equations for the flow with this parameter $k$. An example would be $R_k^{\alpha\beta} = k^2 \delta^{\alpha\beta}$, and typically $R_k^{\alpha\beta}$ is large (has large eigenvalues) for large $k^2$ and vanishes for $k^2\to 0$. In this case the functional probability distribution \eqref{eq:functionalProbDensitySourceCoordinates} approaches a deformed Gaussian for large $k^2$ while for $k^2=0$ it is the standard unmodified probability density of a Euclidean quantum field theory. However, for many steps of the development one can leave the form of $R_k^{\alpha\beta}$ open, and could even consider it as a bilocal source term.

\section{Polchinskis and Wetterichs equations}
As a first step we note the flow equation of the $k$-dependent Schwinger functional,
\begin{equation}
  \frac{\partial}{\partial k} W_k[J] = - \frac{1}{2} \left(\frac{\partial}{\partial k} R_k^{\alpha\beta} \right) \langle \phi_\alpha \phi_\beta \rangle = - \frac{1}{2} \left(\frac{\partial}{\partial k} R_k^{\alpha\beta} \right) \left[ \frac{\delta W_k[J]}{\delta J^\alpha \delta J^\beta} + \frac{\delta W_k[J]}{\delta J^\alpha} \frac{\delta W_k[J]}{\delta J^\beta}\right].
  \label{eq:FlowWk}
\end{equation}
This is the Polchinski equation \cite{Polchinski:1983gv}. It uses the connected correlation function
\begin{equation}
  G_{k\alpha\beta}[J] = \langle \phi_\alpha \phi_\beta \rangle - \langle \phi_\alpha \rangle \langle \phi_\beta \rangle = \frac{\delta^2 W_k[J]}{\delta J^\alpha \delta J^\beta},
  \label{eq:ConnectedCorrelationFunction}
\end{equation}
and the expectation value
\begin{equation}
  \Phi_\alpha[J] = \langle \phi_\alpha \rangle = \frac{\delta W_k[J]}{\delta J^\alpha}.
  \label{eq:expValueSource}
\end{equation}
Incidentally the connected correlation function \eqref{eq:ConnectedCorrelationFunction} is also the Fisher information metric corresponding to the class of probability densities \eqref{eq:functionalProbDensitySourceCoordinates} when sources $J^\alpha$ are taken as coordinates \cite{FloerchingerInformationGeometryEuclidean2023}.

Let us also introduce the Legendre transform
\begin{equation}
  \tilde \Gamma_k[\Phi] = \sup_J \left( J^\alpha \Phi_\beta -W_k[J] \right),
  \label{eq:defGammatilde}
\end{equation}
for which the flow equation is
\begin{equation}
  \frac{\partial}{\partial k} \tilde \Gamma_k[\Phi]{\big |}_\Phi = - \frac{\partial}{\partial k} W_k[J]{\big |}_J = \frac{1}{2} \left( \frac{\partial}{\partial k} R_k^{\alpha\beta} \right) \left[ G_{k\alpha\beta}[\Phi] + \Phi_\alpha \Phi_\beta \right].
  \label{eq:WetterichEqTildeGamma}
\end{equation}
On the right hand side one can write the conncted two-point function as
\begin{equation}
  G_{k\alpha\beta}[\Phi] = ( \tilde \Gamma_k^{(2)}[\Phi] )^{-1}_{\alpha\beta},
\end{equation}
which is the inverse of the second functional derivative of eq.\ \eqref{eq:defGammatilde},
\begin{equation}
  (\tilde\Gamma_k^{(2)}[\Phi])^{\alpha\beta}
  = \frac{\delta^2 }{\delta \Phi_\alpha\delta\Phi_\beta} \tilde \Gamma_k[\Phi].
\label{eq:Gammak2}
\end{equation}
The inverse propagator in \eqref{eq:Gammak2} is actually the Fisher information metric corresponding to \eqref{eq:functionalProbDensitySourceCoordinates} in expectation value coordinates \cite{FloerchingerInformationGeometryEuclidean2023}. 

Eq.\ \eqref{eq:WetterichEqTildeGamma} becomes even nicer in terms of the flowing action
\begin{equation}
  \Gamma_k[\Phi] = \tilde \Gamma_k[\Phi] - \frac{1}{2} R_k^{\alpha\beta} \Phi_\alpha\Phi_\beta,
\end{equation}
which has the flow equation
\begin{equation}
  \frac{\partial}{\partial k} \Gamma_k[\Phi] = \frac{1}{2} \left( \frac{\partial}{\partial k} R_k^{\alpha\beta} \right) ( \Gamma_k^{(2)}[\Phi] + R_k )^{-1}_{\alpha\beta}.
  \label{eq:WetterichEqGamma}
\end{equation}
This is Wetterichs equation \cite{Wetterich:1992yh}.

\section{Flow of the divergence functional}
Consider now the functional Kullback-Leibler divergence between the distributions at source fields $J$ and $J^\prime$, respectively,
\begin{equation}
  \begin{split}
    \tilde D_k[J \| J^\prime] = & \int D\phi \, p_k[\phi, J] \ln(p_k[\phi, J] / p_k[\phi, J^\prime]) \\  
    = & (J^\alpha - J^{\prime\alpha}) \frac{\delta W_k[J]}{\delta J^\alpha} - W_k[J] + W_k[J^\prime],
  \end{split}
\end{equation}
where the first line is the general definition of a Kullback-Leibler divergence and the last line uses the concrete form in eq.\ \eqref{eq:functionalProbDensitySourceCoordinates}. It has the form of a Bregman divergence with reversed arguments.

Alternatively one may replace the sources $J$ and $J^\prime$ by the corresponding expectation values $\Phi$ and $\Phi^\prime$ as coordinates. One finds
\begin{equation}
  \tilde D_k[\Phi \| \Phi^\prime] = \tilde \Gamma_k[\Phi] - \tilde \Gamma_k[\Phi^\prime] - \frac{\delta \tilde \Gamma_k[\Phi^\prime]}{\delta \Phi^\prime_\lambda} (\Phi_\lambda - \Phi^\prime_\lambda).
  \label{eq:DivergenceFunctionalExpectationValueCoordinates}
\end{equation}
It is convenient to also define a divergence functional with subtracted regulator terms
\begin{equation}
  \begin{split}
    D_k[\Phi \| \Phi^\prime] = & \tilde D_k[\Phi \| \Phi^\prime] - \frac{1}{2} R_k^{\alpha\beta} (\Phi_\alpha-\Phi_\alpha^\prime) (\Phi_\beta-\Phi_\beta^\prime) \\
    = & \Gamma_k[\Phi] - \Gamma_k[\Phi^\prime] - \frac{\delta \Gamma_k[\Phi^\prime]}{\delta \Phi^\prime_\lambda} (\Phi_\lambda - \Phi^\prime_\lambda).
  \end{split}
  \label{eq:definitionFlowingDivergence}
\end{equation}
We will call this the flowing divergence. As we will see below it has particularly nice limits. Note, however, that only $\tilde D_k[\Phi \| \Phi^\prime]$ has all the mathematical properties of a relative entropy functional and it is the Bregman divergence associated to $\tilde\Gamma_k[\Phi]$. The flowing divergence \eqref{eq:definitionFlowingDivergence} also vanishes when $\Phi=\Phi^\prime$, but could also be negative for some choices of argument. First derivatives are given by
\begin{equation}
  \begin{split}
    \frac{\delta}{\delta \Phi_\alpha} D_k[\Phi \| \Phi^\prime] = & J^\alpha - J^{\prime\alpha} - R_k^{\alpha\beta}(\Phi_\beta - \Phi_\beta^\prime) = \frac{\delta \Gamma_k[\Phi]}{\delta\Phi_\alpha} - \frac{\delta \Gamma_k[\Phi^\prime]}{\delta\Phi^\prime_\alpha},  \\
    \frac{\delta}{\delta \Phi_\alpha^\prime} D_k[\Phi \| \Phi^\prime] = & - \frac{\delta^2 \Gamma_k[\Phi^\prime]}{\delta \Phi_\alpha^\prime \delta\Phi^\prime_\lambda} (\Phi_\lambda - \Phi^\prime_\lambda).
  \end{split}
\end{equation}

One may easily work out the second functional derivatives
\begin{equation}
  \begin{split}
    (\tilde D_k^{(2,0)}[\Phi \| \Phi^\prime])^{\alpha\beta} = \frac{\delta^2}{\delta\Phi_\alpha\delta\Phi_\beta} \tilde D_k[\Phi \| \Phi^\prime] =  & (\tilde \Gamma_k^{(2)}[\Phi])^{\alpha\beta}, \\
    (\tilde D_k^{(1,1)}[\Phi \| \Phi^\prime])^{\alpha\beta} = \frac{\delta^2}{\delta\Phi_\alpha\delta\Phi_\beta^\prime} \tilde D_k[\Phi \| \Phi^\prime] = & - (\tilde \Gamma_k[\Phi^\prime])^{\alpha\beta}, \\
    (\tilde D_k^{(0,2)}[\Phi \| \Phi^\prime])^{\alpha\beta} = \frac{\delta^2}{\delta\Phi_\alpha^\prime\delta\Phi_\beta^\prime} \tilde D_k[\Phi \| \Phi^\prime] = & - (\Phi_\lambda - \Phi^\prime_\lambda) \frac{\delta^3}{\delta\Phi^\prime_\alpha\delta\Phi^\prime_\beta \delta\Phi^\prime_\lambda} \tilde \Gamma_k[\Phi^\prime]  + (\tilde \Gamma_k[\Phi^\prime])^{\alpha\beta}.
  \end{split}
  \label{eq:secondFunctionalDerivativesTildeD}
\end{equation}
It is interesting to note that the right hand side of the first line depends actually only on $\Phi$. Higher order functional derivatives give one-particle irreducible vertex functions in the presence of the regulator $R_k$.  In a similar way, the right hand side of the second line depends only on $\Phi^\prime$. Finally, the term on the right hand side of the third line is linear in $\Phi$ but depends non-linearly on $\Phi^\prime$.

From eq.\ \eqref{eq:WetterichEqTildeGamma}, together with its derivative with respect to the field expectation value argument, we find a flow equation for the divergence functional
\begin{equation}
  \begin{split}
  \frac{\partial}{\partial k} \tilde D_k[\Phi \| \Phi^\prime] = \frac{1}{2} \left( \frac{\partial}{\partial k} R_k^{\alpha\beta} \right) {\bigg [} & (\Phi_\alpha - \Phi^\prime_\alpha)(\Phi_\beta - \Phi_\beta^\prime) + (\tilde \Gamma_k^{(2)}[\Phi])^{-1}_{\alpha\beta} - (\tilde \Gamma_k^{(2)}[\Phi^\prime])^{-1}_{\alpha\beta}  \\
  & +  (\tilde \Gamma_k^{(2)}[\Phi^\prime])^{-1}_{\alpha\mu} (\tilde \Gamma_k^{(2)}[\Phi^\prime])^{-1}_{\beta\nu}  (\Phi_\lambda - \Phi_\lambda^\prime) \frac{\delta^3 \tilde\Gamma_k[\Phi^\prime]}{\delta\Phi^\prime_\mu \delta \Phi_\nu^\prime \delta\Phi^\prime_\lambda}
  {\bigg ]}.
  \end{split}
\end{equation}
Here one may use \eqref{eq:secondFunctionalDerivativesTildeD} which yields a closed flow equation for the modified divergence functional,
\begin{equation}
  \begin{split}
  \frac{\partial}{\partial k} \tilde D_k[\Phi \| \Phi^\prime] = \frac{1}{2} \left( \frac{\partial}{\partial k} R_k^{\alpha\beta} \right) {\bigg [} & 
  (\Phi_\alpha - \Phi^\prime_\alpha)(\Phi_\beta - \Phi_\beta^\prime) + 
  (\tilde D_k^{(2,0)}[\Phi \| {\color{lightgray}\Phi^\prime}])^{-1}_{\alpha\beta}  \\
  & -  (\tilde D_k^{(1,1)}[{\color{lightgray}\Phi} \| \Phi^\prime])^{-1}_{\alpha\lambda}   \; 
  (\tilde D_k^{(0,2)}[\Phi \| \Phi^\prime])^{\lambda\kappa} \;
  (\tilde D_k^{(1,1)}[{\color{lightgray}\Phi} \| \Phi^\prime])^{-1}_{\kappa\beta}  
  {\bigg ]}.
  \end{split}
\end{equation}
On the right hand side we have greyed out those arguments on which the correspondonding functionals do not depend any more after the functional derivative have been taken. Finally, let us write this in terms of the flowing divergence,
\begin{equation}
  \begin{split}
  \frac{\partial}{\partial k} D_k[\Phi \| \Phi^\prime] = \frac{1}{2} & \left( \frac{\partial}{\partial k} R_k^{\alpha\beta} \right) {\bigg [} 
  (D_k^{(2,0)}[\Phi \| {\color{lightgray}\Phi^\prime}]+R_k)^{-1}_{\alpha\beta}  \\
  & -  (D_k^{(1,1)}[{\color{lightgray}\Phi} \| \Phi^\prime]+R_k)^{-1}_{\alpha\lambda}   \; 
  (D_k^{(0,2)}[\Phi \| \Phi^\prime] + R_k)^{\lambda\kappa} \;
  (D_k^{(1,1)}[{\color{lightgray}\Phi} \| \Phi^\prime] + R_k)^{-1}_{\kappa\beta}  
  {\bigg ]}.
  \end{split}
  \label{eq:flowEquation}
\end{equation}
This exact flow equation is our main result.

\section{Limits of large and vanishing regulator}

From the definitions in \eqref{eq:definitionFlowingDivergence}, \eqref{eq:defGammatilde} and \eqref{eq:modifiedSchwingerFunctional} one can obtain the functional integral relation for the flowing divergence,
\begin{equation}
  e^{-D_k[\Phi \| \Phi^\prime]} = \frac{\int D\phi \, \exp\left( -S[\phi] - \frac{1}{2}R_k^{\alpha\beta}(\phi_\alpha-\Phi_\alpha)(\phi_\beta-\Phi_\beta) + \frac{\delta}{\delta \Phi_\alpha} \Gamma_k[\Phi] (\phi_\alpha - \Phi_\alpha) \right)}{\int D\tilde \phi \, \exp\left( -S[\tilde \phi] - \frac{1}{2}R_k^{\alpha\beta}(\tilde\phi_\alpha-\Phi_\alpha^\prime)(\tilde\phi_\beta-\Phi_\beta^\prime) + \frac{\delta}{\delta\Phi_\alpha^\prime} \Gamma_k[\Phi^\prime]  (\tilde\phi_\alpha-\Phi_\alpha) \right)}.
  \label{eq:DivergenceFunctionalIntegralExpectationValuesOnly}
\end{equation}
One could further replace here
\begin{equation}
  \begin{split}
    \frac{\delta}{\delta\Phi_\alpha} \Gamma_k[\Phi] = & \frac{\delta}{\delta\Phi_\alpha} D_k[\Phi \| \Phi_k^\text{eq}], \\
    \frac{\delta}{\delta\Phi_\alpha^\prime} \Gamma_k[\Phi^\prime] = & - \frac{\delta}{\delta\Phi_\alpha} D_k[\Phi \| \Phi^\prime]{\big |}_{\Phi = \Phi_k^\text{eq}},
  \end{split}
  \label{eq:DeterminationJFromSphiphi}
\end{equation}
where $\Phi_k^\text{eq}$ is the expectation value configuration solving $\delta\Gamma_k[\Phi]/\delta\Phi_\alpha = 0$. 

For large regulator scale $k$, where $R_k^{\alpha\beta} \sim k^2 \delta^{\alpha\beta}$ is assumed, fluctuations around the expectation values are suppressed and a saddle point approximation becomes valid
\begin{equation}
  \lim_{k\to \infty}D_k[\Phi \| \Phi^\prime] =  S[\Phi] - S[\Phi^\prime] - \frac{\delta}{\delta \Phi^\prime_\alpha} S[\Phi^\prime] (\Phi_\alpha - \Phi^\prime_\alpha).
  \label{eq:SteepestDescend02}
\end{equation}
This can be supplemented by the next-to-leading order which is a one-loop term in the presence of the regulator $R_k$. The crucial feature of eq.\ \eqref{eq:SteepestDescend02} is that the right hand side is known! It is fully specified by the microscopic action $S[\phi]$ entering the probability density \eqref{eq:functionalProbDensitySourceCoordinates}.

In the opposite limit the flowing divergence approaches the full Kullback-Leibler divergence or relative entropy functional,
\begin{equation}
  \lim_{k\to 0}D_k[\Phi \| \Phi^\prime] = D[\Phi \| \Phi^\prime].
\end{equation}
The physical significance of $D[\Phi \| \Phi^\prime]$ is discussed in ref.\ \cite{FloerchingerInformationGeometryEuclidean2023}.

We find thus that $D[\Phi \| \Phi^\prime]$ is fully determined by the starting point \eqref{eq:SteepestDescend02} and the solution to the flow equation \eqref{eq:flowEquation}.

\section{Conclusions}
We have derived here a renormalization group flow equation for the flowing divergence functional, which is a generalization of the Kullback-Leibler divergence to the setup of Euclidean quantum field theories or statistical field theories. The new flow equation is a close relative of Wetterich's flow equation for the flowing action \cite{Wetterich:1992yh}.

The flowing action is defined such that it approaches for large regulator an expression fully defined by the microscopic action $S[\phi]$ or, in other words, the probability density at vanishing source $J$. In the opposite limit of a vanishing regulator the flowing divergence equals the full functional Kullback-Leibler divergence. The latter can be seen as a generating functional for correlation functions and it has an information theoretic significance expressed for example through Sanov's theorem \cite{FloerchingerInformationGeometryEuclidean2023}.

The flow equation we derived here is a functional differential equation for the steps between the two limiting cases. It can therefore be used to determine the divergence functional through the solution of a renormalization group flow instead of a perturbative or Monte-Carlo calculation, for example. This makes new approximation schemed possible, very much as in other applications of the functional renormalization group in quantum or statistical field theory.

Future generalizations of the setup discussed here concern in particular quantum relative entropies and dynamical situations in quantum field theories. Relative entropies have the advantage that they are well defined in the context of local quantum field theories \cite{arakiRelativeEntropyStates1977}, in contrast to von-Neumann entanglement entropies which suffer from ultraviolet divergences for reduced states corresponding to bounded regions of space.

Our new flow equation contributes also to the growing field of information geometry and could find applications beyond physics.

\section*{Acknowledgement}
This work is supported by the Deutsche Forschungsgemeinschaft (DFG, German Research Foundation) under 273811115 – SFB 1225 ISOQUANT.

\providecommand{\href}[2]{#2}\begingroup\raggedright\endgroup

\end{document}